# Magnetic properties of carbon phases synthesized using high pressure-high temperature treatment.

K.-H. Han[*,1], A. Talyzin[1], A. Dzwilewski[1], T. L. Makarova[1],
R. Höhne[2], P. Esquinazi[2], D. Spemann[2], and L. S. Dubrovinsky[3]

[1] Department of Physics, Umeå University, S-90187, Umeå, Sweden
[2] Superconductivity and Magnetism Division, University of Leipzig, D-04103, Leipzig, Germany
[3] Bayerisches Geoinstitut, Universität Bayreuth, D-95440 Bayreuth, Germany.

**Abstract.**

Two sets of samples were synthesized at 3.5 GPa near the point of $C_{60}$ cage collapse at different annealing times. A clear structural transformation from mixture of $C_{60}$ polymeric phases to graphite-like hard carbon phase was confirmed by X-ray diffraction and Raman spectroscopy. Magnetic force microscopy and superconducting quantum interference device were used to characterize the magnetic properties of the synthesized samples. We found that the sample preparation conditions used in this study are not suitable to produce bulk magnetic carbon.

---

[*] Email: kyoohyun.han@physics.umu.se

## 1. Introduction

Peculiar magnetic properties have been reported for high pressure-high temperature (HPHT) polymerized fullerenes.[1-7] Ferromagnetic behavior at near room temperature has been reported for samples synthesized from $C_{60}$ by HPHT treatment at 2.5 GPa,[5-7] 6 GPa,[1,3] and 9 GPa.[2,4] Magnetic properties of these samples were studied using superconducting quantum interference device (SQUID)[1-3,6] and magnetic force microscopy (MFM).[6,7] However, some of the samples studied in these papers were later discovered to show contamination with magnetic impurities on the levels much higher than initially reported,[1,6-9] while for some others the level of contamination was not reported.[2,3] Contamination with hydrogen was also reported in recent study for one sample prepared at 9 GPa.[4] Contrary to the initial report, it also appeared that similar ferromagnetic properties were observed for samples with very different structure: not only polymeric $C_{60}$ but also graphite-like carbon which is formed as a result of fullerene molecules collapse.[7,8]

One of the main supporting arguments for existence of intrinsic ferromagnetism of carbon has been obtained using MFM method. Using MFM magnetic domain patterns in several kinds of carbon materials have been studied like photopolymerized fullerenes,[10] pressure-polymerized $C_{60}$ fullerenes,[6,7] proton irradiated micro-spots[11] and large spots[12] on highly oriented pyrolytic graphite (HOPG) samples. Only one sample synthesized at HPHT conditions (at 2.5 GPa and 1123 K) has been studied by MFM in previous publications[6,7] (In Ref. 6 the preparation conditions for the same sample have been erroneously reported as 6 GPa and 1123 K). The sample synthesized at such conditions is most likely composed of graphitic hard carbon.[5] Detailed MFM study of this sample revealed some regions with a magnetic domain structure, which has been considered as an evidence for the intrinsic nature of the observed ferromagnetism.

However, direct assignment of the observed bulk ferromagnetism with the domain features was difficult due to high contamination with magnetic impurities.[6,9] Thus, the amount of carbon contribution to the bulk ferromagnetism remained an open question. New experiments with less contaminated samples are required in order to reveal the true contribution of magnetic domain areas to bulk ferromagnetism.

For this study two sets of samples were synthesized at 3.5 GPa in a broad temperature range 923-1193 K (below and above the point of $C_{60}$ collapse) and different annealing time while extra care was taken to hold the contamination with ferromagnetic impurities during synthesis and handling of the samples as low as possible. The structure of each sample was studied by X-ray diffraction (XRD) and Raman spectroscopy. We used MFM as a tool to study magnetic microstructures, while the bulk magnetization was measured using SQUID magnetometer. Here, we present a study of magnetic domain patterns in correlation with structural modifications of samples produced from $C_{60}$ by HPHT treatment.

**2. Experimental**

For the synthesis, we used high purity, twice-sublimed $C_{60}$ powder with a nominal amount of iron impurity of ~22 μg/g (~5 ppm) provided by Term Corporation (USA) as a starting material. All operations with the samples before and after synthesis were performed using plastic instruments.

The samples were prepared using standard piston-cylinder system with the piston size of ½ inch. An advantage of our piston-cylinder method is relatively large size of samples (150-170 mg), but maximal pressure is only 3.5-4 GPa. All materials used for preparation of high pressure assemblage were of high purity in respect to metallic impurities in order to avoid contaminations during the synthesis. Pristine $C_{60}$ powder was loaded into a platinum or gold capsule with tight cover. The capsule was placed into a standard assemblage, which consists of pyrex and alumina tubes while the resistive annealing was provided using high purity graphite tube. We prepared two sets of samples at constant pressure of 3.5 GPa. The first set of samples was synthesized at constant annealing time of 1000 sec at different temperatures (T = 923-1193 K) and the second set of samples was synthesized at constant temperature (1085 K) at different annealing times (60-10800 sec).[13] The HPHT treatment was performed using following steps:

1. Annealing for 5-8 minutes at 0.5 GPa and 450 K.

2. Slow increase of pressure to desired value (3.5 GPa).

3. Rapid annealing to target temperature 923 K ≤ T ≤ 1193 K for the first set of samples (1085 K for the second set of samples)

4. Annealing at target temperature during 1000 sec for the first set of samples (60 sec to 10800 sec for the second set of samples).

5. Quenching with the cooling rate of about 80 degree/sec under constant pressure.

6. After cooling down to 300 K the pressure was lowered slowly to atmospheric pressure.

Sample number, detailed synthesis conditions, the impurity concentrations as well as saturation magnetization measured with the SQUID at 5 K and 300 K of the second set of samples are shown in Table 1. Notice that sample No. 4 in Table I belongs to both set of samples. The second set of samples showed that the point at which fullerene molecules collapse also depends on the annealing time. Samples of $C_{60}$ polymeric phases were obtained in the form of powder together with larger pieces (several mm sizes). The samples of graphite-like carbon were prepared as cylinders with a layered texture. Due to this texture, it was easy to split the initial cylinders into sets of discs.

Structural characterization of the samples was performed using a Renishaw Raman 2000 spectrometer equipped with 514 nm laser with a resolution of 2 cm$^{-1}$. Powder X-ray diffraction patterns were recorded using a Philips system based on a PW1820 goniometer.

In all samples magnetic force gradient images and sample's topography were obtained simultaneously with a Nanoscope IV scanning probe microscope from Digital Instruments at room temperature. The microscope was operated in the "tapping/lift $^{TM}$" scanning mode with phase detection because this mode allows the separation of the topographic effect from the magnetic signal. All the images were taken with standard magnetic tip (Type MESP) for various tip-sample distances (50 to 300 nm). Other details of MFM measurement can be found elsewhere.[6,7,10] The second set of samples was studied for bulk magnetization using SQUID magnetometer from Quantum Design with the reciprocating sample option and a sensitivity of ~$10^{-10}$ Am$^2$.

The purity of samples studied by SQUID was analyzed using Particle Induced X-ray Emission (PIXE) method.[6,7] This method provides high accuracy for magnetic impurities; however the analysis is limited to the depth of ~40 μm. It is possible that surface contamination could be different from bulk contamination and results of this analysis should be treated with some caution. Analysis of impurity levels by PIXE revealed that the main magnetic impurity is iron and it is distributed non-uniformly in the sample: for two samples (No. 5 and 6 in Table I) the average level was about 28-30 μg/g while one of the samples (No. 4) was found to be more contaminated (up to 94 μg/g). Other magnetic impurities were found on the level below 3-4 μg/g and a relatively high contamination with Si was also observed for all samples (90-120 μg/g).

## 3. Results.

### 3.1 Structural characterization using XRD and Raman spectroscopy.

Figure 1(a) shows the X-ray diffraction patterns of the first set of samples, synthesized at 3.5 GPa for 1000 sec at different temperatures. This figure proves that in the temperature region 923-1073 K the synthesized samples consist of a mixture of $C_{60}$ polymeric phases. The pattern shown for 923 K sample is interpreted as mostly rhombohedral (Rh) polymeric phase, while increase of temperature to 953 K results in appearance of tetragonal (T) polymeric phase. The amount of tetragonal polymeric phase increases with temperature up to 1073 K.

According to P-T diagram of $C_{60}$ at HPHT conditions available from literature,[1,5] fullerene molecules collapse if heated above some critical temperature $T_c$. In our first set of samples a clear phase transformation is observed between 1073 K and 1085 K in very good agreement with previous publications.[1] This phase transformation is known from literature and explained by the collapse of $C_{60}$ cage structure with formation of so called "graphite-like hard carbon" phase.[14-15] This name was given to the carbon phase which typically exhibits high hardness (up to ~30 GPa[14]) and poor crystallinity. The XRD patterns of these samples show only one broad peak, which corresponds to the (002) reflection of graphite. It should be noted that the intensity of this peak was rather low and the pattern was scaled up for Figure 1.

Figure 1(b) shows the XRD patterns of the second set of samples (synthesized at 3.5 GPa, 1085 K with different annealing times 60-10800 sec). The samples with shorter annealing times appeared to be of polymeric $C_{60}$ structure while three samples with longer annealing times (1000-10800 sec) exhibited typical "hard carbon" phase, similar to the patterns shown in Fig. 1(a) for the first set of samples at T > 1085 K.

The data obtained by XRD were also confirmed by Raman spectroscopy. Sharp peaks due to Rh and T-polymers of $C_{60}$ are replaced by very broad features typical for graphite-like amorphous carbon starting from 1085 K. It should be noted that the relative intensity of the Raman peaks from T-polymer increases as the samples annealing temperature increases from 923 to 1073 K, which is in good agreement with XRD data. The pressure of 3.5 GPa is smaller compared to pressure of 6 GPa which has been most often used for synthesis of Rh-polymers previously. This explains why many of our samples are composed by a mixture of both Rh and T-polymers. More detail structural characterization of these samples will be given elsewhere.[13]

**3.2 Results of MFM**

All samples were examined by MFM. MFM images were recorded in randomly chosen areas of sample surface (about 30-50 scans). We grouped the samples according to the synthesis conditions and magnetic domain patterns:

- *Group A*: Samples prepared at 923 K ≤ T ≤ 983 K and at T = 1093 K and samples prepared at annealing times 60–480 sec (No. 1-3 in Table I). Magnetic domain patterns without stripes, which are correlated with topography, were found in some areas of samples.

- *Group B*: Samples prepared at 1085 K with annealing times of 1000 sec (No. 4 in Table I) and 3600 sec (No. 5 in Table I). Magnetic domain patterns with parallel stripes were found in small areas (~ 10%) of the whole sample surface. The structure of these samples was graphite-like.

- *Group C*: Samples prepared at 1053 K and 1073 K (first set of samples). No magnetic images were found in the whole sample surface. The structure of the samples was polymeric $C_{60}$.

- *Group D*: Two samples synthesized at T = 1193 K and at T = 1085 K with 10800 sec (No. 6 in Table I) annealing time. No magnetic images were found in the whole area of the samples. The structure of the samples was graphite-like.

The relation of groups A-D to synthesis conditions and structures of samples is summarized in Figure 2.

Figure 3 shows topographic (left) and magnetic force gradient (right) images obtained in samples from *Group A* (923 K(a), 953 K(b) and 983 K(c)). Even though the change of phase shift is small ($|\Delta\Phi| \leq 0.2°$), we could resolve magnetic force gradient images. The shape of magnetic force gradient images is cluster-like, similar to the topography and it reveals that the magnetic images are highly correlated to the topography. Images of this kind were also obtained in samples No 1-3. The average root-mean-square value of surface roughness ($S_q$) in magnetic areas is ~50 nm and all the MFM images were obtained with the scan height, h ≥ 50 nm. We couldn't resolve any magnetic images in other areas. In some cases high roughness could result in artifacts on magnetic images, but in our experiments no magnetic images were obtained even for some regions where roughness was significantly higher (~90 nm). Thus, we argue that the magnetic images are not an artifact, which may come from the sample topography, but an intrinsic effect of sample.

Large changes in the magnetic domain configuration were found in samples of the *Group B*, however the roughness of topography remains similar to that of other samples. Figure 4 shows topographic (left) and magnetic force gradient (right) images of samples in *Group B* (No. 4 ((a) and (b)) and No. 5 (c)). Two kinds of magnetic force gradient images were found in these samples: cluster-like shape (a) and stripe lines ((b) and (c)). Cluster-like magnetic patterns are correlated with topographic images (Fig. 4(a)) and the change of phase shift is small ($|\Delta\Phi| \leq 0.2°$). Clear stripe magnetic domain patterns (Figs. 4(b) and 4(c)) were observed and the maximum change of phase shift is $|\Delta\Phi| \approx 1°$. Strongly aligned stripes were observed in sample No. 5 (Fig. 4(c)) and these domain patterns indicate the presence of a much more collective magnetic behavior than that found in the samples of shorter and longer annealing time. The sharply bright and dark images in MFM represent a significant out of plane component of the magnetization.[16] The stripes are parallel to each other and the width and length of domain patterns are ~ 1 μm and ~ 10 μm, respectively.

Figures 5(a) and 5(b) show topographic (left) and magnetic force gradient (right) images of samples in *Group C* (1053 K) and *Group D* (No. 6), respectively. The average size of topographic clusters and surface roughness are similar to other samples however, we couldn't resolve any magnetic domain patterns in a whole area of sample surface.

In order to estimate the value of the magnetization at the stripe area in samples in Group B (No 4 and 5), we measured MFM with different scan heights and chose one position where the value of phase shift is maximum, as shown in Figures 6(a) and 6(b) by the arrows. We used the point dipole approximation [17,18] and assumed that the direction of magnetization at the chosen point is perpendicular to the surface ($z$ direction) with a magnetization $M_0$. Following the same procedure as in Ref. 17, we obtained a magnetization $M_0$ = 14.5 Am$^2$/kg (16.8-29.5 Am$^2$/kg) with R =180.7 nm (165.4 nm) and t = 1000 nm for sample No. 4 (No. 5), where *R* is the spot radius given by the 1/*e* decay length of the magnetic stray field of the spot at a given position and t is the effective thickness of the magnetic area. For comparison, this value (16.8-29.5 $Am^2/kg$) is similar to those of large area proton irradiated HOPG (21.4 $Am^2/kg$) [12] and polymerized $C_{60}$ (~10 $Am^2/kg$).[6]

### 3.3 Magnetization measurements using SQUID

Several samples were characterized by SQUID magnetometer at 5 K and 300 K. Results of these measurements are shown in Figure 7. Special attention was paid to two samples prepared at T = 1085 K with t = 1000 sec (No. 4) and 3600 sec (No. 5), where most pronounced domain structures were found by MFM. As it can be seen in Fig. 7(a) all samples exhibited clearly diamagnetic behavior. After subtraction of the diamagnetic background very weak ferromagnetic loops were obtained with saturation magnetization of the order of $10^{-4}$ $Am^2/kg$. Figure 7(b) shows ferromagnetic loops for all studied samples, besides sample 2. Somewhat stronger saturation magnetization and somewhat more regular ferromagnetic loops were found for two samples which were slightly contaminated during the synthesis: these samples exhibited cracked gold capsule after HPHT experiment (sample No. 1 and 3, Table I). Saturation magnetization values are given in the Table I. The data shown in this table clearly demonstrate that the saturation magnetization values do not correlate with the structure of samples, MFM data and level of impurities determined from PIXE. Similarly small magnetization was found for both the samples with $C_{60}$ polymeric structure and for samples where fullerene collapsed into graphite-like hard carbon phase. No increase in saturation magnetization relative to the other samples was found in samples No. 4 and 5, where MFM shows parallel stripe magnetic images.

It should be noted that saturation magnetization on the level of $1 \times 10^{-4}$ $Am^2/kg$ could be explained by concentration of Fe impurity of only 0.5 μg/g assuming that all that iron behaves ferromagnetically. The concentration of Fe impurities in our samples was measured up to an average level of ~100 µg/g. The increase of impurity levels compared to pristine $C_{60}$ powder (~22 µg/g) is likely due to contamination introduced during the HPHT synthesis. One of the samples (No. 4) of graphite-like carbon (synthesized at 3.5 GPa, 1085 K and 1000 sec annealing time) was studied by PIXE in 5 different regions of ~0.8 mm diameter. The Fe impurity was found on the level of 70-210 µg/g, clearly different in different areas, averaging at ~94 µg/g. Amounts of Ni and Co were determined in this sample as 5.2 µg/g and 0.5 µg/g respectively. The results showed us that the impurity is distributed not uniformly on the surface of samples. Therefore, the value of contamination given in the Table I must be considered only as approximate. No clear correlation was found between amounts of magnetic impurities determined from PIXE and values of saturation magnetization, while slightly increased $M_s$ for samples with cracked gold capsules points out to the contribution of a higher contamination in these samples.

It can be concluded that measurements of the bulk magnetization, after subtraction of the diamagnetic background, shows weak ferromagnetic-like signals in all samples studied by SQUID, besides No. 2. The enhanced diamagnetic behavior of some samples can be understood taking into account their graphite-like structure found from XRD. Interestingly, diamagnetic properties were also found for all studied samples of polymeric $C_{60}$, independently from the ratio between Rh and T phases. This observation contradicts previous results [2] which showed that paramagnetic response of pressure-polymerized fullerenes changes to a diamagnetic response of graphite-like carbon above the point of fullerene collapse.

### 3.4 Discussion.

The experimental MFM results presented above can be summarized shortly as follows:

- MFM do not exclude some magnetic structures in samples with different atomic structures: $C_{60}$ polymeric and graphite-like "hard carbon" phases. In most of these samples the domains appear to be correlated to the topography.

- Most pronounced domain structures (covered at least 10% of the surface in one of the samples) were observed on samples with completely collapsed fullerene molecules. Only on the material with the structure of graphite-like hard carbon exhibited magnetic stripe domains, which were not correlated with the topography.

Although domain structures found in some samples were similar to the previously reported,[6] the samples produced in this work showed a bulk ferromagnetism with magnetization at saturation on the level of $10^{-4}$ Am$^2$/kg, i.e. much smaller than in previously reported samples.[1,9] These saturation values showed neither correlation to MFM data nor to the impurity concentrations. We stress, however, that the value of magnetization is obtained dividing the measured magnetic moment by the total sample mass. The real value of the magnetization of the magnetically ordered material is certainly much larger since most of the sample is not ferromagnetic, as MFM measurements indicate. Note that values of the order of $10^{-4}$ Am$^2$/kg could be explained by ferromagnetically ordered Fe impurity on the level of ~ 0.5-3 µg/g while the measured Fe contamination was up to 30 times higher.

The magnetic domain structures observed in this study are not correlated to different amount of magnetic impurities. These magnetic structures suggest the existence of intrinsic magnetism in these samples. The same starting powder and experimental routines were used in all experiments and it is difficult to explain the observed variation of magnetic properties by different amount of impurities. In principle, if some fine dispersed impurity was present in the starting powder, it could undergo clusterization at the moment of $C_{60}$ collapse. Some sharp changes in the magnetic properties could result from such a process. In our experiments the sharpest MFM pictures were, in fact, observed on the sample synthesized right above the collapse point of $C_{60}$ but we suggest that the observed domain images are connected to a change in the sample structure or sample surface properties rather than impurity redistribution.

MFM observations of domain structures in graphite-like "hard carbon" presented in this study seem to be in agreement with models of defect-induced magnetization.[4] The crystal structure of samples synthesized right above the point of fullerene collapse is certainly strongly defective, possibly some graphene sheets are curved and some dangling bonds are present. The open question is about the magnitude of the ferromagnetism due to defects in carbon structures as well as the influence of light atoms, like hydrogen, oxygen or nitrogen during the synthesization process. Recent theoretical results emphasize the possible influence of hydrogen on the magnetism of graphite-like structures.[19,20]

Regarding the difference between MFM data and bulk magnetization measurements we note that:

- The overall amount of domain areas was about 2-3 times smaller compared to previously studied sample,[2,3] while the difference in the bulk magnetization at saturation was 2-3 orders of magnitude.

- ferromagnetic domain structures are likely to be limited only to the surface of material, for example due to defects created as a result of surface reconstruction

In summary, $C_{60}$ samples with low amount of magnetic impurities were synthesized by HPHT treatment at various annealing temperatures and annealing times, close to fullerene collapse point (both below and above). We can conclude that in this study neither $C_{60}$ polymer phases nor the samples where fullerene cages collapse into graphite-like hard carbon phases indicated the existence of bulk ferromagnetism, unlike to previously reported data obtained for samples prepared under different pressure conditions. The reasons for the observation of magnetism on a surface micro-level should be clarified in future studies.

**Acknowledgements**

The work done in Sweden was supported by the Swedish Research Council, RFBR 05-02-17779 and the FP6 program "Ferrocarbon". The studies in Leipzig were supported by the FP6 program "Ferrocarbon" and the Deutsche Forschungsgemeinschaft under DFG ES 86/11-1. A. Dzwilewski thanks Helge Ax:son Johnson foundation for support. We thank G. D. Bromiley for help with sample preparation. Large scale facility program is acknowledged for providing access to high pressure equipment.

**Table caption**

Table I. The table shows the sample numbers, synthesis conditions, Fe and Ni impurity concentrations and the saturation magnetization at two temperatures obtained after the subtraction of the diamagnetic contribution.

n.d. : not determined.

**Figure captions**

Figure 1. X-ray diffraction patterns from samples synthesized at 3.5 GPa and 1000 sec annealing time with different annealing temperature (a) and from samples synthesized at 3.5 GPa and 1085 K using different annealing times (b). Arrows show two peaks from tetragonal polymer, which change their relative intensity depending on annealing temperature.

Figure 2. Schematic representation of MFM results relative to their synthesis conditions and structure. Symbols correspond to following sample groups (see text): ■-group A (cluster-shape patterns),▲-group B (parallel stripes), ×-group C (no patterns), +- group D (no patterns).

Figure 3. Topographic (left) and magnetic force gradient (right) images obtained from the samples of group A the samples prepared at 923 K (a), 953 K (b), and 983 K (c). Here scan size is 2 μm × 1 μm (a and b) and 2 μm × 0.5 μm (c) and scan height is 50 nm (a and c) and 100 nm (b) for MFM.

Figure 4. Topographic (left) and magnetic force gradient(right) images for samples of group B  prepared at 1085 K and annealing times of 1000 sec (a and b)  and 3600 sec

(c). Here scan size is 3 μm × 3 μm (a), 2 μm × 2 μm (b) and 5 μm × 3.75 μm (c) and scan height is 50 nm for MFM.

Figure 5. Topographic (left) and magnetic force gradient (right) images for samples of group C (at 1053 K (a)) and group D (at 1085 K and 10800 sec (b)). Here scan size is 2 μm × 1 μm (a), 2 μm × 2 μm (b) and scan height of MFM is 50 nm for all samples.

Figure 6. Scan height dependence of phase shift for samples No. 4 (a) and No. 5 (b)

Figure 7. (a) Magnetization data of six samples measured by SQUID magnetometer (numbers correspond to those in Table 1), (b) Ferromagnetic loops obtained after subtraction of diamagnetic background.

| No. | Synthesis conditions (3.5 GPa) | Fe, µg/g | Ni, µg/g | $M_s$, Am$^2$/kg (300K) | $M_s$, Am$^2$/kg (5K) |
|---|---|---|---|---|---|
| 1 | 1085K, 60s | n.d. | n.d. | 0.00014 | 0.00014 |
| 2 | 1085K, 180s | 58 | 4.0 | 0 | 0 |
| 3 | 1085K, 480s | 43 | 3.7 | 0.00032 | 0.00032 |
| 4 | 1085K, 1000s | 94 | 5.2 | 0.0001 | 0.0001 |
| 5 | 1085K, 3600s | 30 | 3.0 | 0.00006 | 0.00007 |
| 6 | 1085K, 10800s | 28 | 3.4 | 0.00013 | 0.00014 |

Table I.

Figure 1 by K. –H. Han et al.

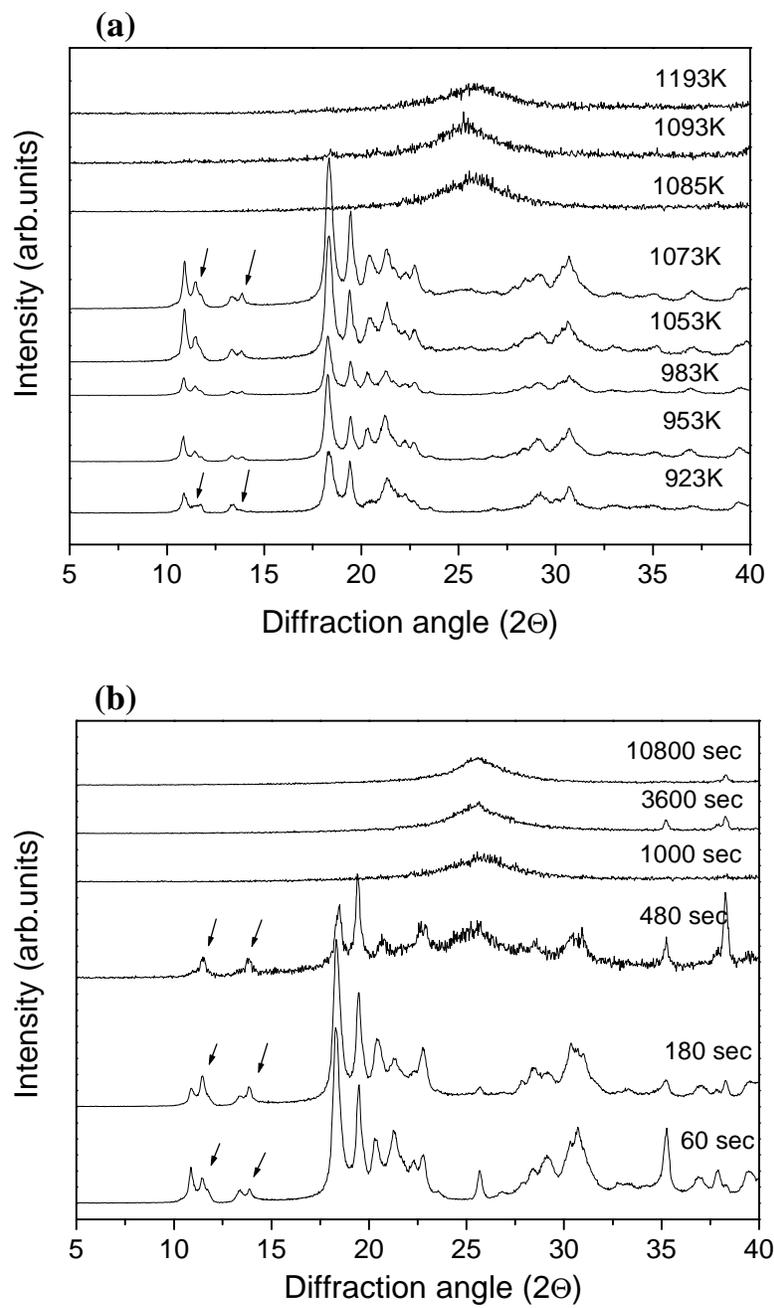

Figure 2 by K. –H. Han et al.

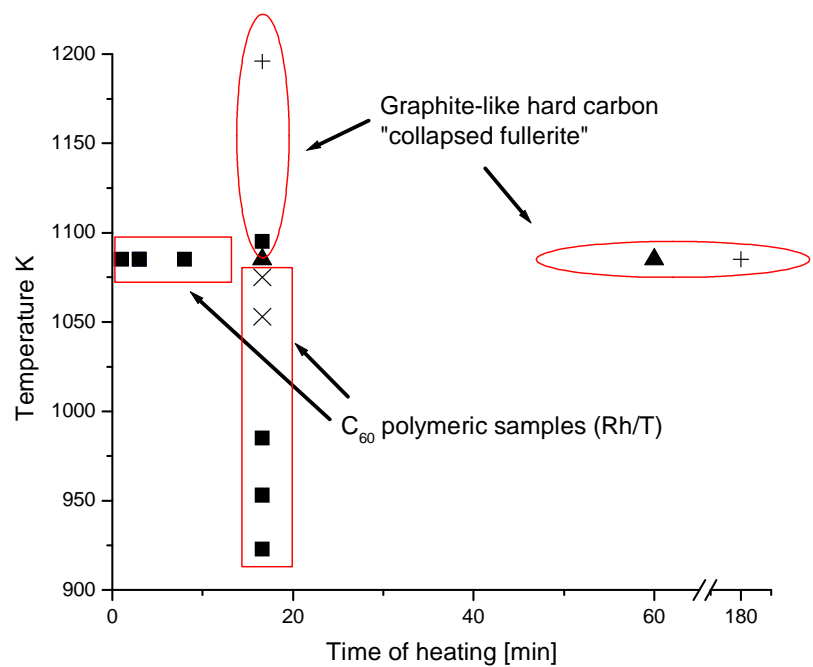

Figure 3 by K. –H. Han et al.

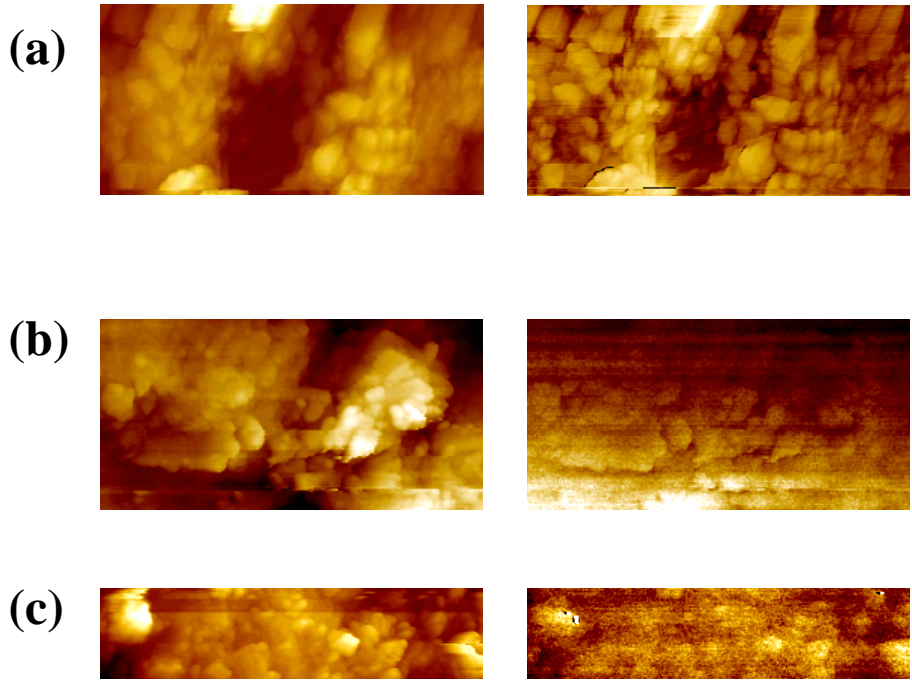

Figure 4 by K. –H. Han et al.

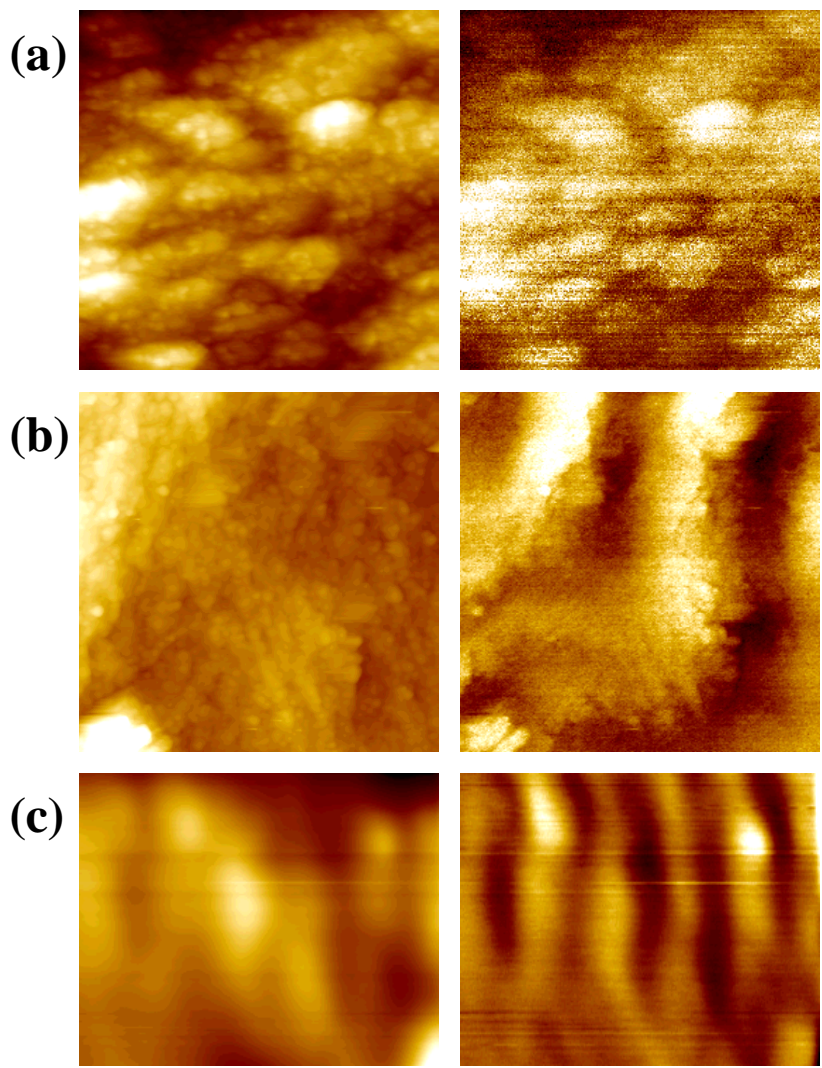

Figure 5 by K. –H. Han et al.

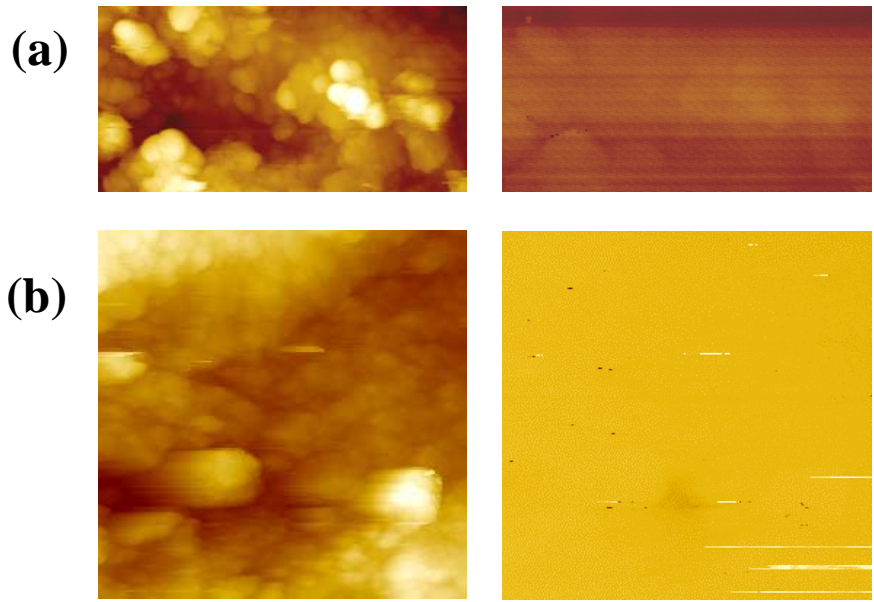

Figure 6 by K. –H. Han et al.

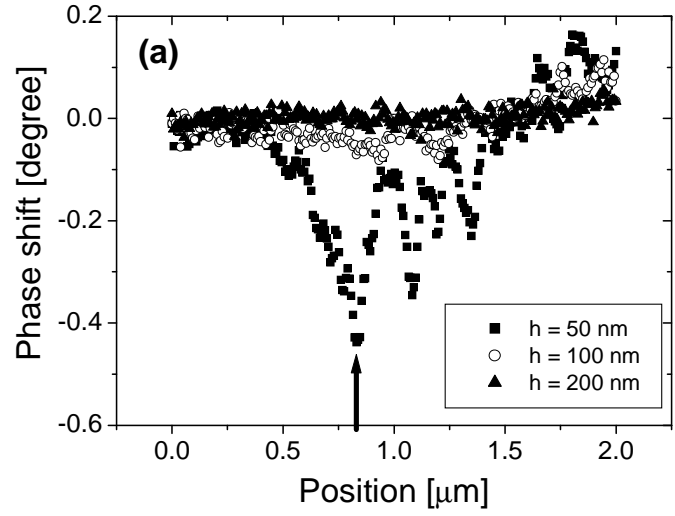

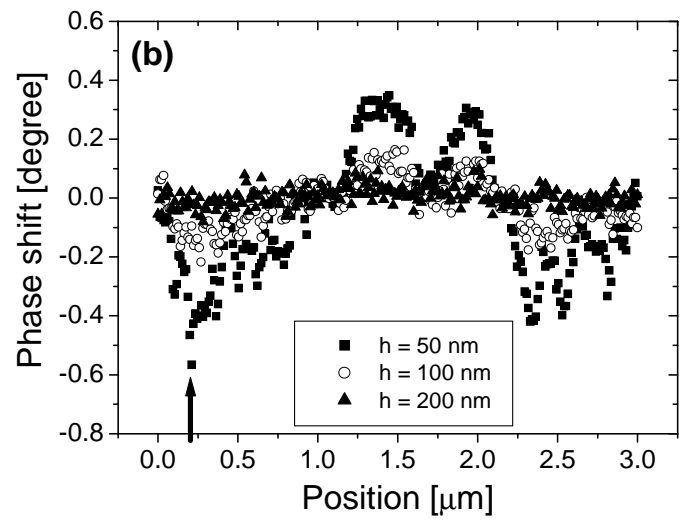

Figure 7 by K. –H. Han et al.

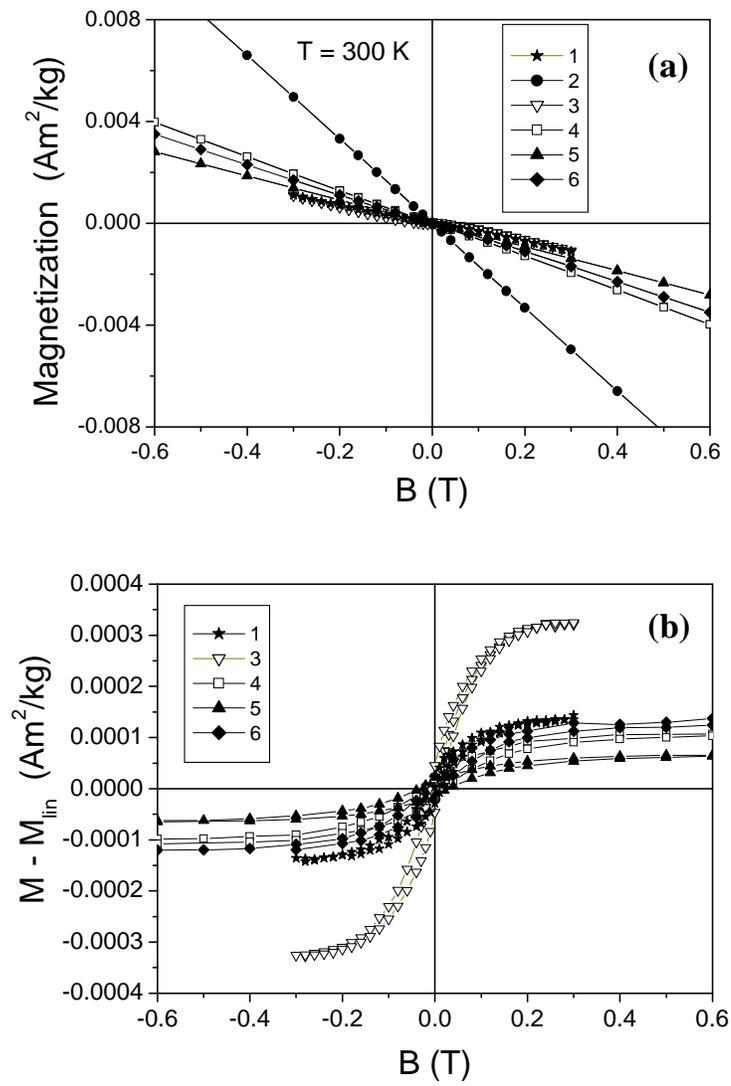